\documentclass[twocolumn]{aastex631}

\usepackage[T1]{fontenc}

\shorttitle{}
\shortauthors{Hasegawa et al.}
\graphicspath{{./}{figures/}}

\begin{document}

\title{Bulk and atmospheric metallicities as direct probes of sequentially varying accretion mechanisms of gas and solids onto planets}

\author[0000-0002-9017-3663]{Yasuhiro Hasegawa}
\affiliation{Jet Propulsion Laboratory, California Institute of Technology, Pasadena, CA 91109, USA}
\email{yasuhiro.hasegawa@jpl.nasa.gov}

\author[0000-0002-0919-4468]{Mark R. Swain}
\affiliation{Jet Propulsion Laboratory, California Institute of Technology, Pasadena, CA 91109, USA}



\begin{abstract}

Core accretion is the standard scenario of planet formation,
wherein planets are formed by sequential accretion of gas and solids, 
and is widely used to interpret exoplanet observations.
However, no direct probes of the scenario have been discussed yet.
Here, we introduce an onion-like model as one idealization of sequential accretion and propose that bulk and atmospheric metallicities of exoplanets can be used as direct probes of the process.
Our analytical calculations, coupled with observational data, demonstrate that the trend of observed exoplanets supports the sequential accretion hypothesis. 
In particular, accretion of planetesimals that are $\gtrsim $ 100 km in size is most favored to consistently explain the observed trends.
The importance of opening gaps in both planetesimal and gas disks following planetary growth is also identified.
New classification is proposed, wherein most observed planets are classified into two interior statuses: globally mixed and locally (well-)mixed.
Explicit identification of the locally (well-)mixed status enables reliable verification of sequential accretion.
During the JWST era, the quality and volume of observational data will increase drastically and improve exoplanet characterization.
This work provides one key reference of how both the bulk and atmospheric metallicities can be used to constrain gas and solid accretion mechanisms of planets.

\end{abstract}



\section{Introduction} \label{sec:intro}

Uncovering the ubiquity of planetary systems around main-sequence stars is a fundamental finding of contemporary astrophysics 
\citep[e.g.,][]{1995Natur.378..355M,2011arXiv1109.2497M,2011ApJ...728..117B,2018ApJS..235...38T}.
Planets discovered orbiting other stars, so-called extrasolar planets or exoplanets in short, exhibit profound diversity in their properties including mass, orbital periods, and multiplicity.
Such diversity challenges the canonical theory of planet formation that was developed solely from the solar system.
Hence, exoplanets provide a novel opportunity of testing our understanding of planet formation {\it statistically} \citep[e.g.,][]{2004ApJ...604..388I,2009A&A...501.1139M,2014prpl.conf..691B}.

\begin{figure*}
\centering
\includegraphics[width=1\linewidth]{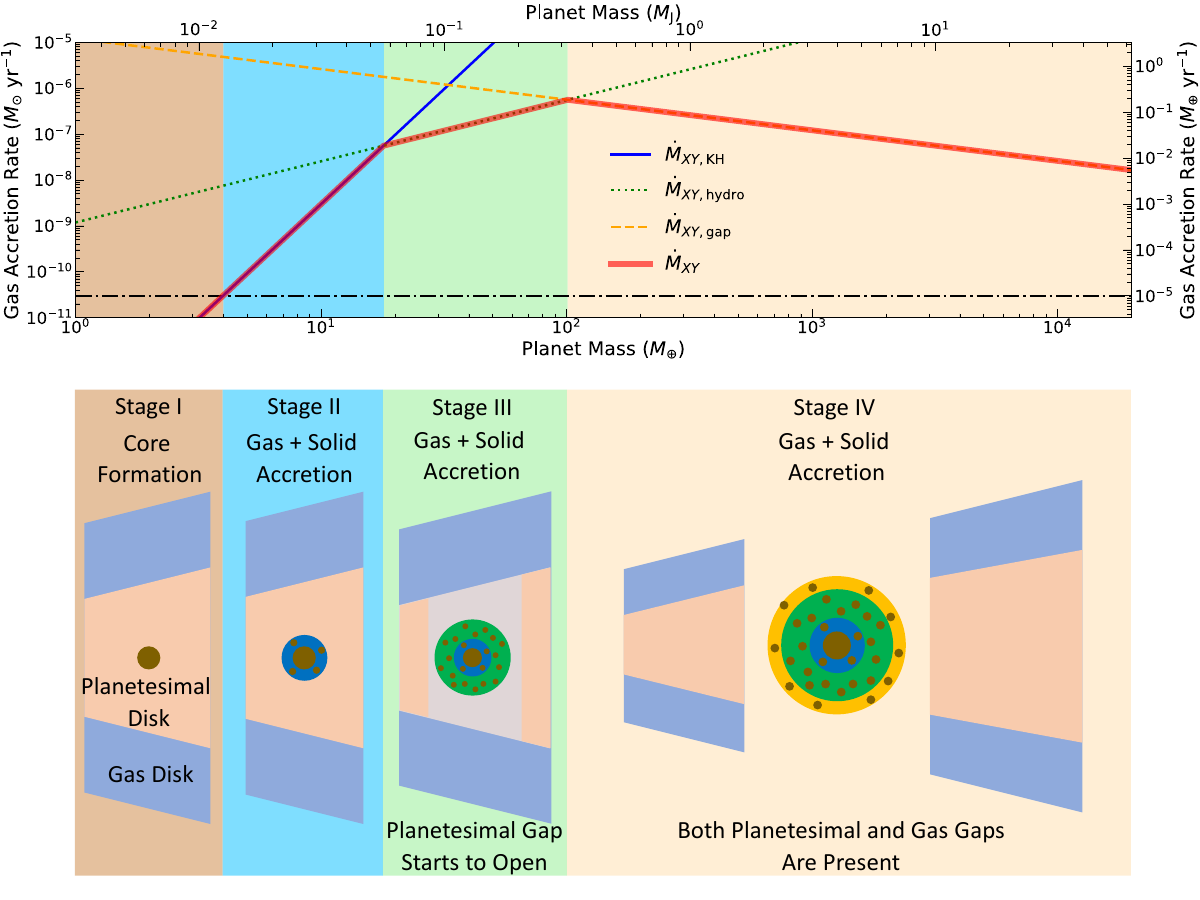}
\caption{Four stages of planet formation via core accretion and its schematic diagram.
{\it Top:} Sequential accretion of gas and solid divides the mass growth of planets into four stages.
In Stage I, planetary cores form by solid accretion. 
As time goes on, solid accretion rates decrease, and gas accretion becomes dominated.
One conservative estimate of a characteristic solid accretion rate is $10 M_{\oplus}/1$ Myr $\sim 10^{-5} M_{\oplus}$ yr$^{-1}$ as denoted by the black dash-dotted line.
Once the solid accretion rate becomes lower the value, gas accretion comes into play.
In Stage II, accretion of both gas and solids occurs. The gas accretion rate is regulated by cooling of accreted gas (i.e., $\dot{M}_{XY, \rm KH}$ denoted by the blue solid line).
In Stage III, gas accretion is controlled by disk evolution. The corresponding rate becomes very high (i.e., $\dot{M}_{XY, \rm hydro}$ denoted by the green dotted line), 
and it is the main channel of forming sub-giants.
In Stage IV, gas accretion is reduced due to planet-disk interaction and the resulting gap formation in gas disks (i.e., $\dot{M}_{XY, \rm gap}$ denoted by the yellow dashed line).
This is the final stage of giant plane formation.
Mathematical expressions of accretion rates are summarized in Appendix \ref{sec:app_model}.
{\it Bottom:} Sequential accretion of gas and solid can be idealized by the onion-like model.
As we show below, bulk and atmospheric metallicities can be used to trace when gap formation in both planetesimal and gas disks occurs,
which corresponds to Stages III and IV, respectively.
The purple and light salmon trapezoids represent gas and planetesimals disks, respectively. 
The brown circles denotes solids and the circular rings represent envelope gas.}
\label{fig1}
\end{figure*}

It is widely recognized that the standard theory of planet formation is core accretion \citep[e.g.,][]{1996Icar..124...62P}.
This scenario considers that growing (proto)planets undergo sequential accretion of gas and solids, wherein a dominant mode varies with increasing planet mass.
The first step is to build planetary cores through accretion of pebble-sized and/or $\gtrsim$ km-sized solids 
\citep[e.g.,][]{1989Icar...77..330W,1998Icar..131..171K,2010A&A...520A..43O,2012A&A...544A..32L}.
The latter solids are regarded as the parent bodies of asteroids and comets in the solar system and referred to as planetesimals in general.
Once (proto)planets become massive enough ($\sim 5-10 M_{\oplus}$, where $M_{\oplus}$ is the Earth mass), 
then they proceed to the second step in which they start accreting surrounding gas simultaneously with solids \citep[e.g.,][]{1980PThPh..64..544M,1982P&SS...30..755S}.
The gas accretion stage is known to divide into several sub-stages \citep[e.g.,][]{2000Icar..143....2B,2012A&A...547A.111M,2019ApJ...876L..32H};
combining the core formation stage, the four-stage sequence is summarized in Figure \ref{fig1}.

Currently, the above sequential accretion of gas and solids is the central basis to understand the origin of solar system planets as well as exoplanets.
The validity of the scenario has been implied from various {\it in}direct evidence such as the presence of various kinds of planets;
different accretion stages may correspond to paths of forming different kinds of planets.
However, no {\it direct} probes of such sequential accretion have been discussed in the literature yet.

We here propose that the combination of exoplanet bulk and atmospheric metallicities, for a spectrum of exoplanet masses, can be used as a direct probe of the sequential accretion model. 
Different mass ranges probe the different stages in the sequential accretion scenario. 
Theoretical models can be used to make specific predictions about the relationship between mass and metallicity for the different stages of the sequential accretion scenario. 
Existing observational data are utilized to perform a proof-of-concept study.
In light of surge of characterization of exoplanet atmospheres,
this work provides one key reference to better understand the origin of exoplanets and their diversity.

\section{Core accretion and bulk and envelope metallicities }

\subsection{Basic picture}

The standard picture of core accretion naturally supports the presence of three stages in gas accretion 
\citep[e.g.,][see Figure \ref{fig1}]{1996Icar..124...62P,2000Icar..143....2B,2012A&A...547A.111M}.
Mathematical expressions of accretion rates are summarized in Appendix \ref{sec:app_model}.

Initially, (proto)planets gain surrounding gas nearly spherical-symmetrically (Stage II in Figure \ref{fig1}).
In this stage, gravitational capture and the subsequent cooling of the accreted gas determine the growth rate of planets.
The corresponding timescale is often called the Kelvin-Helmholtz timescale.
The gas accretion rate ($\dot{M}_{XY, \rm KH}$) in Stage II accelerates rapidly with planet mass,
and hence the resulting accretion is regulated eventually by supply of surrounding gas and switched to a different mode of gas accretion.
The mode is referred to as the disk-limited gas accretion (Stage III in Figure \ref{fig1}) 
and is one main channel of forming sub-Saturn planets due to high accretion rates ($\dot{M}_{XY, \rm hydro}$).
In this stage, growing planets are  massive ($\gtrsim 10 M_{\oplus}$) enough to interact gravitationally with the surrounding gas,
and as a result, angular momentum between planets and the disk gas is transferred, leading to changes in the gas density around the planets.
More massive ($\gtrsim 100 M_{\oplus}$) planets considerably affects the surrounding gas abundance and finally open up a gap in the spatial distribution of gas.
Gap formation in gas disks is a signal of the onset of the final gas accretion stage (Stage IV in Figure \ref{fig1}),
where the resulting accretion rate ($\dot{M}_{XY, \rm gap}$) is reduced due to the presence of a gas gap.

Accompanying solid accretion determines the bulk and atmospheric metallicities of planets.
Suppose that a planet has the gas and solid masses of $M_{XY}$ and $M_Z$, respectively,
with the total mass ($M_{\rm p}$) of
\begin{equation}
M_{\rm p}  =  M_{XY} + M_{Z}.
\end{equation}
The solid mass can be further decomposed into two components:
\begin{equation}
M_{Z}  =  M_{Z, \rm core} + M_{Z, \rm env},
\end{equation}
where $M_{Z, \rm core}$ and $M_{Z, \rm env}$ are the solid masses constituting a planetary core and distributing in a gaseous envelope, respectively.
Then, bulk ($Z_{\rm p, bulk}$) and envelope ($Z_{\rm p, env}$) metallicities of the planet are defined as
\begin{eqnarray}
\label{eq:Z_bulk}
Z_{\rm p, bulk} & \equiv &  \frac{ M_{Z} }{ M_{\rm p} },  \\ 
\label{eq:Z_env}
Z_{\rm p, env}  & \equiv & \frac{ M_{Z, \rm env} }{ M_{XY} +  M_{Z, \rm env}}.   
\end{eqnarray}
We use atmospheric and envelope metallicities interchangeably.

\subsection{Onion-like model}

Examining the sequential accretion picture can be facilitated by introducing an onion-like model (Figure \ref{fig1}).

The model is one idealization of sequential accretion of gas and solids and based on two assumptions:
The first one is that heavy elements accreted during the $i$th stage are well mixed with the gas accreted during the same stage.
This situation could be achieved by thermal ablation that occurs when accreted solids enter and pass through planetary envelopes.
The assumption would be reasonable, especially if accreted solids are volatile-rich or not too large in size.
The second assumption is that the resulting envelope metallicity is maintained even at subsequent gas accretion stages.
While the validity of this assumption is much more questionable than the first one,
it might be possible for some volatiles.

Under the onion-like model, planets thereby establish a layer of envelope metallicities as they move to different gas accretion stages,
which can be written as 
\begin{eqnarray}
Z_{\rm p, env}^{i} & \equiv  & \left.  \frac{ \Delta M_{Z, \rm env} }{ \Delta M_{XY} +  \Delta M_{Z, \rm env}} \right|_{i} \\ \nonumber
                                               & =          &  \left.  \frac{ \dot{M}_{Z, \rm env} \Delta t}{ ( \dot{M}_{XY} +  \dot{M}_{Z, \rm env} ) \Delta t} \right|_{i} 
                                                \simeq    \left .\frac{ \dot{M}_{Z, \rm env}  }{ \dot{M}_{XY} }   \right|_{i}           
\end{eqnarray}
where $\Delta M_{XY}$ and $\Delta M_{Z, \rm env}$ are the gas and solid masses accreted during the $i$th stage, respectively,
and $\Delta t$ is the time span of the $i$th stage.
The corresponding gas and solid accretion rates are expressed by $\dot{M}_{XY} $ and $ \dot{M}_{Z, \rm env}$, respectively.
The resulting $Z_{\rm p, env}^{i} $ should be viewed as {\it primordial} envelope metallicities.
It is reasonable to consider that $ \dot{M}_{Z, \rm env}  \ll \dot{M}_{XY} $ at Stages II-IV; 
their envelopes are composed mainly of gas.

\subsection{Theoretical prediction} \label{sec:pred}

There are three characteristic solids that are important for metal enrichment of planets \citep{2018ApJ...865...32H}.
These solids are classified by size: dust, pebbles, and planetesimals.
As shown in \citet{2018ApJ...865...32H} and confirmed in Section \ref{sec:comp},
planetesimal accretion provides a most consistent explanation of the total heavy-element mass trend of the observed exoplanets (also see Appendix \ref{sec:app_model}).
For this case, the presence or absence of gaps in planetesimal disks around planets becomes one key parameter (Figure \ref{fig1}).
We therefore focus on planetesimal accretion and explore how planetesimal accretion can reproduce the atmospheric metallicity trend, 
paying attention to the presence or absence of planetesimal gaps.

{\it Stage I:} If formed planets underwent only core formation, then the primordial planet properties are
\begin{eqnarray}
M_{\rm p}                  & \sim & M_Z \sim M_{Z, \rm core}, \\ \nonumber
M_{Z, \rm env} & \sim & 0.
\end{eqnarray}
Equivalently, from equations (\ref{eq:Z_bulk}) and (\ref{eq:Z_env}),
\begin{eqnarray}
\label{eq:Zp_StI}
Z_{\rm p, bulk}^{\rm StI} & \simeq & 1, \\ \nonumber
Z_{\rm p, env}^{\rm StI}  & \simeq  & 0.
\end{eqnarray}
Stage I corresponds to the so-called superEarth regime,
and planets in the regime could be formed by other processes such as giant impacts.
Their atmospheres would be affected significantly by outgassing or suffer readily from evaporation if they are in the vicinity of the host star.
Accordingly, this stage is covered only for completeness purpose.

{\it Stage II:} In this stage, gas accretion gets just started, 
and it is reasonable to assume that $M_{\rm p} \sim M_{Z, \rm core} \gtrsim M_{XY}$ and $M_Z \sim M_{Z, \rm core} \gg M_{Z, \rm env}$.
Then, from equation (\ref{eq:Z_bulk}), 
\begin{equation}
\label{eq:Zbulk_StII}
Z_{\rm p, bulk}^{\rm StII} \simeq 1.
\end{equation}
On the other hand, the envelope metallicity is computed as, under the onion-like model,
\begin{eqnarray}
\label{eq:Zenv_StII}
Z_{\rm p, env}^{\rm StII}  & \simeq &   \left .\frac{ \dot{M}_{Z, \rm env}  }{ \dot{M}_{XY} }   \right|_{\rm StII}  = \frac{ \dot{M}_{Z, \rm nogap}  }{ \dot{M}_{XY, \rm KH} }   \\ \nonumber  
                                  & \simeq &4.5 \times 10^{-2}  \left( \frac{f_{\rm grain}}{ 10^{-3} } \right)^{1/5} \left( \frac{\Sigma_{\rm s}}{2.7 \mbox{ g cm}^{-2}}  \right)    \\ \nonumber     
                                    & \times & \left( \frac{r_{\rm p}}{5 \mbox{ au}} \right)^{6/5}   \left( \frac{M_{\rm p}}{4M_{\oplus}} \right)^{-7/5},                                                                                                                                                                                                                                                                           
\end{eqnarray}
where the case of no planetesimal gap is adopted (i.e., $\dot{M}_{Z, \rm env}=\dot{M}_{Z, \rm nogap}$),
$f_{\rm grain} (\ll 1)$ is a parameter taking into account the effect of reduction in grain opacity of planetary envelopes, 
and $f_{\rm grain}=10^{-3}$ is adopted to better reproduce the population of observed exoplanets \citep{2014A&A...566A.141M,2014ApJ...794...25H},
$ \Sigma_{\rm s}$ is the solid surface density, and $r_{\rm p}$ is the position of planets.
In the above equation, the so-called minimum mass solar nebula (MMSN) model is used for reference \citep{1981PThPS..70...35H}.

{\it Stage III:} This is the main stage of forming sub-giants. 
Previous studies show that $M_{\rm p} \sim M_{XY}$ and $M_{Z} \simeq M_{Z, \rm env}(\ga 10 M_{\oplus})$ for observed giant exoplanets \citep{2016ApJ...831...64T}.
It is important to recognize that given that one conservative estimate of the critical core mass is $10 M_{\oplus}$,
the latter finding confirms that 
solid accretion after core formation is most critical for determining the bulk metallicities of these planets \citep{2018ApJ...865...32H}.
Accordingly, the bulk metallicity of planets is given as
\begin{equation}
Z_{\rm p, bulk}^{\rm StIII} \simeq \frac{M_{Z, \rm env}}{M_{XY}} \simeq  \left .\frac{ \dot{M}_{Z, \rm env}  \Delta t}{ \dot{M}_{XY} \Delta t}   \right|_{\rm StIII} .
\end{equation}
As we show in Section \ref{sec:comp}, both the cases with and without planetesimal gaps are crucial for reproducing the trend of observed exoplanets.
Consequently, bulk metallicities are written as, for the case of no gap formation (i.e., $\dot{M}_{Z, \rm env}=\dot{M}_{Z, \rm nogap}$),
\begin{eqnarray}
\label{eq:Zbulk_no_StIII}
Z_{\rm p, bulk}^{\rm StIII} & \simeq & \frac{ \dot{M}_{Z, \rm nogap} }{ \dot{M}_{XY, \rm hydro} } \\ \nonumber
                                        & \simeq &6.7 \times 10^{-3}   \\ \nonumber    
                                        & \times  & \left( \frac{\Sigma_{\rm s}}{2.7 \mbox{ g cm}^{-2}}  \right)  \left( \frac{\Sigma_{\rm g}}{1.5 \times 10^{-2} \mbox{ g cm}^{-2}}  \right)^{-1/4}   \\ \nonumber    
                                        & \times &   \left( \frac{H_{\rm g}/r_{\rm p}}{0.05} \right)^{2/5}  \left( \frac{r_{\rm p}}{5 \mbox{ au}} \right)^{11/10} \left( \frac{M_{\rm p}}{20 M_{\oplus}} \right)^{-8/15},
 \end{eqnarray}
and, for the case of gap formation (i.e., $\dot{M}_{Z, \rm env}=\dot{M}_{Z, \rm gap}$)
\begin{eqnarray}
\label{eq:Zbulk_yes_StIII}
Z_{\rm p, bulk}^{\rm StIII}  & \simeq & \frac{ \dot{M}_{Z, \rm gap} }{ \dot{M}_{XY, \rm hydro} } \\ \nonumber
                                          &  \simeq &    7.3 \times 10^{-3}  \left( \frac{\tau_{\rm damp}}{10^4 \mbox{ yr}} \right)^{7/10}  \\ \nonumber      
                                          & \times &  \left( \frac{\Sigma_{\rm s}}{2.7 \mbox{ g cm}^{-2} }   \right)  \left( \frac{\Sigma_{\rm g}}{1.5 \times 10^2 \mbox{ g cm}^{-2}}  \right)^{2/5} \\ \nonumber  
                                          & \times &    \left( \frac{H_{\rm g}/r_{\rm p}}{0.05} \right)^{4/5} \left( \frac{r_{\rm p}}{5 \mbox{ au}} \right)^{5/4}    \left( \frac{M_{\rm p}}{20 M_{\oplus}} \right)^{-3/10} ,
\end{eqnarray}
where $\Sigma_{\rm g}$ and $H_{\rm g}$ are the surface density and pressure scale height of the surrounding gas, respectively, 
and $\tau_{\rm damp}$ is the timescale of damping the eccentricity of planetesimals.
Again, the MMSN model is used to compute reference values of disk properties.

Under the onion-like model, the envelope metallicities are given as 
\begin{equation}
\label{eq:Zenv_StIII}
Z_{\rm p, env}^{\rm StIII}  \simeq   \left. \frac{ \dot{M}_{Z, \rm env}  }{ \dot{M}_{XY} }   \right|_{\rm StIII} = \frac{ \dot{M}_{Z, \rm env}  }{ \dot{M}_{XY, \rm hydro} } \simeq Z_{\rm p, bulk}^{\rm StIII}. 
\end{equation}

{\it Stage IV:} This is the final stage of giant planet formation.
As described above, previous studies show that  $M_{\rm p} \sim M_{XY}$ and $M_{Z} \simeq M_{Z, \rm env} (\ga 10 M_{\oplus})$ \citep{2016ApJ...831...64T},
suggesting that most of heavy elements should be accreted during Stage III \citep{2018ApJ...865...32H}.
Hence 
\begin{equation}
\label{eq:Zbulk_StIV}
Z_{\rm p, bulk}^{\rm StIV}  \simeq \frac{ \dot{M}_{Z, \rm env}  }{ \dot{M}_{XY, \rm hydro} } \simeq Z_{\rm p, bulk}^{\rm StIII}. 
\end{equation}
For envelope metallicities, planetesimal accretion with gap formation better reproduces the trend of observed exoplanets, as shown in Section \ref{sec:comp}.
Then, they are written as
\begin{eqnarray}
\label{eq:Zenv_StIV}
Z_{\rm p, env}^{\rm StIV}  & \simeq & \frac{ \dot{M}_{Z, \rm gap} }{ \dot{M}_{XY, \rm gap} } \\ \nonumber
                                          & \simeq & 3.8 \times 10^{-4}  \left( \frac{\tau_{\rm damp}}{10^4 \mbox{ yr}} \right)^{7/10}   \left( \frac{\alpha}{10^{-2}}  \right)^{2/5} \\ \nonumber
                                           & \times &  \left( \frac{\Sigma_{\rm s}}{2.7 \mbox{ g cm}^{-2} }   \right)  \left( \frac{\Sigma_{\rm g}}{1.5 \times 10^2 \mbox{ g cm}^{-2}}  \right)^{2/5}  \\ \nonumber      
                                           & \times  &  \left( \frac{H_{\rm g}/r_{\rm p}}{0.05} \right)^{6/5} \left( \frac{r_{\rm p}}{5 \mbox{ au}} \right)^{5/4}   \left( \frac{M_{\rm p}}{100 M_{\oplus}} \right)^{-7/5}  , 
 \end{eqnarray}
where $\alpha$ represents the strength of gas turbulence \citep{1973A&A....24..337S}, 
and one conservative value (i.e., $\alpha=10^{-2}$) that is required to reproduce the observed disk accretion onto pre-main sequence stars \citep{1998ApJ...495..385H}, is adopted above.

\subsection{Comparison} \label{sec:comp}

\begin{figure}
\centering
\includegraphics[width=0.9\linewidth]{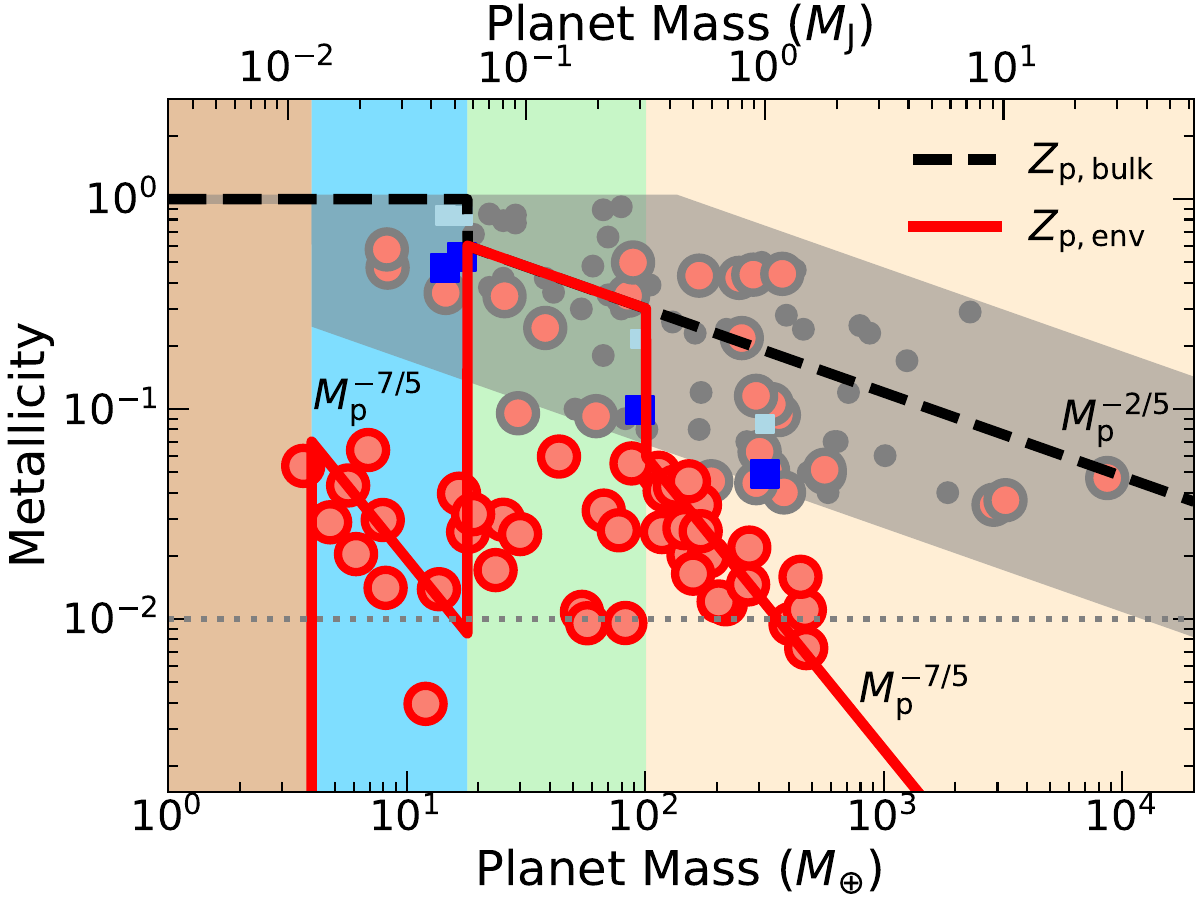}
\caption{Bulk and envelope metallicities of observed exoplanets and their trends.
Bulk and envelope metallicities are denoted by the gray and salmon dots.
For the latter, high and low values are separated by the gray and red edges.
The gray shaded region covers the distribution of the bulk metallicities of observed exoplanets.
The predicted profiles of the bulk and envelope metallicities are denoted by the black dashed and red solid lines, respectively.
To obtain a better fit, the amplitudes of the profiles are adjusted.
For comparison purpose, the solar system planets (i.e., Jupiter, Saturn, Uranus and Neptune) are included \cite[references herein]{2018ApJ...865...32H}.
Their bulk and atmospheric metallicities are denoted by the light and dark blue squares, respectively.}
\label{fig2}
\end{figure}

The above theoretical predictions can be verified, by comparing with the exoplanet data available in the literature (Figure \ref{fig2}).
In contrast to the common approach adopted in previous work \citep[e.g.,][]{2014ApJ...793L..27K},
the mass-based metallicity (not the abundance-based one) is used (Appendix \ref{sec:app_data}).
The mass-based metallicity is preferred when comparing with the outcome of the onion-like model;
while the envelope metallicity predicted from the onion-like model would represent the primordial one at the formation stage,
the observational data are taken from the present-day atmosphere of mature exoplanets.
In order to make a better comparison, 
we have considered all the elements that eventually turn into volatile and refractory materials (except for hydrogen and helium)
when computing the mass-based metallicity.
This implies that the computed mass-based metallicities would represent the {\it primordial} ones under the well-mixed assumption.

The comparison consists of two parts: the absolute value and the trend as a function of planet mass.
Since the absolute value depends sensitively on the initial condition (e.g., solid surface density and the formation timing of planets)
as well as the formation location \citep{2016ApJ...832...41M},
the trend as a function of planet mass may enable more reliable comparison between theory and observations \citep{2016ApJ...832...41M,2018ApJ...865...32H}.
We therefore focus on the latter and discuss implications derived from the former.
The current observational data cover the planet mass of $M_{\rm p} \gtrsim 18 M_{\oplus}$ (i.e., Stages III to IV) for bulk metallicities
and that of $M_{\rm p} \gtrsim 4 M_{\oplus}$ (i.e., Stages II to IV) for envelope metallicities.

We find that combination of planetesimal accretion both without and with gaps (i.e., $(-8/15-3/10)/2 \simeq -0.42$; equations (\ref{eq:Zbulk_no_StIII}) and (\ref{eq:Zbulk_yes_StIII})) 
can better reproduce the profile ($-0.45 \pm0.09$) of the bulk metallicities of observed exoplanets inferred by \citet{2016ApJ...831...64T}.
This is similar to the finding of previous studies \citep{2018ApJ...865...32H}; 
a quantitative difference is caused by explicit consideration of the mass-radius relation when computing the solid accretion rate.
The importance of both the accretion modes indicates that gap formation in planetesimal disks occurs during Stage III on average.
The current calculations (i.e., equations (\ref{eq:Zbulk_no_StIII}) and (\ref{eq:Zbulk_yes_StIII})) predict much lower bulk metallicities,
suggesting that movement of planet-forming materials, either migration of accreting planets or radial drift of pebbles, is required.
For the latter, pebbles arriving at the feeding zone of accreting planets should be converted to planetesimals before being accreted onto the planets, 
which is possible due to various instabilities assisted by local accumulation of pebbles.

For envelope metallicities, our calculations provide reasonable explanations to some of observed exoplanets (Figure \ref{fig2});
in Stages II and IV, the trends of exoplanets that have the envelope metallicity of $\lesssim 0.1$ can be reproduced well,
while in Stage III, our calculated profile is applicable to exoplanets with high envelope metallicities.
Slight enhancement of solids relative to the MMSN model would be sufficient to obtain a better fit for Stage II (equation (\ref{eq:Zenv_StII})).
On the other hand, at least two order of magnitude increase is needed for Stages III and IV (equations (\ref{eq:Zenv_StIII}) and (\ref{eq:Zenv_StIV})).
As described above, movement of planet-forming materials is necessary to better reproduce the current observational data.

In summary, the trend of observed exoplanets supports sequential accretion of gas and solids 
and infers the process of gap formation in planetesimal disks, following planetary growth;
in Stage II, no gap is present in planetesimals disks.
As planets become more massive and move to Stage III, planetesimal gaps are open around the planets.
In Stage IV, planet formation proceeds under the presence of both planetesimal and gas gaps.
Also, movement of planet-forming materials is one essential process to better understand metal enrichment of observed exoplanets.

\section{Discussion}

\begin{figure*}
\centering
\includegraphics[width=0.9\linewidth]{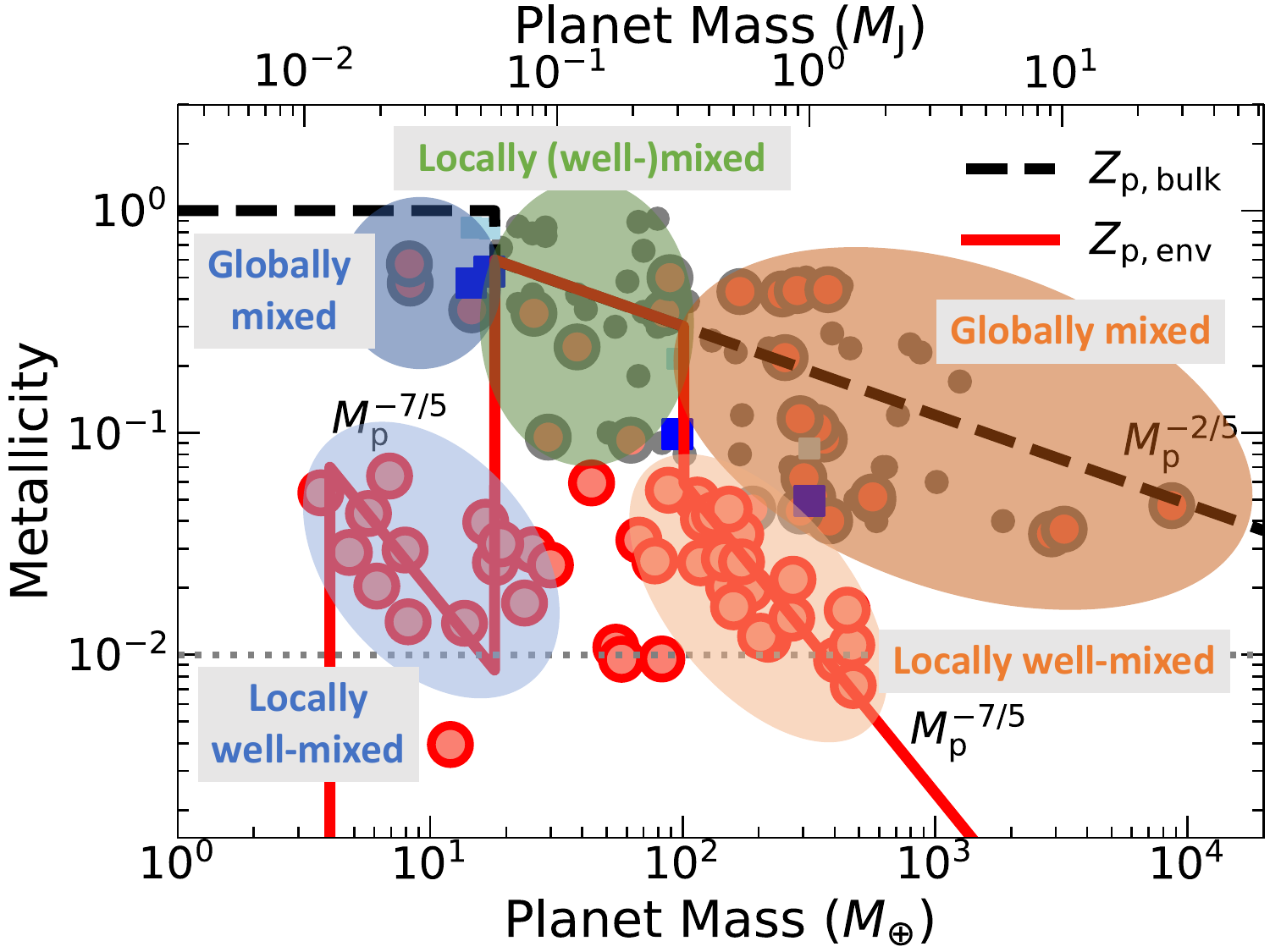}
\caption{Exoplanet classification proposed by the onion-like model.
Most observed planets are classified into two interior statues with some outliers: the globally mixed status and the locally (well-)mixed status.
The shaded ellipses are added only for visualization purpose without quantitive estimates.
Identification of the locally (well-)mixed status enables reliable verification of sequential accretion of gas and solids onto planets.}
\label{fig3}
\end{figure*}

Revealing the origin of the bulk and envelope metallicities of observed exoplanets leads to better characterization.
Based on the above comparison, the following classification can be proposed (Figure \ref{fig3}):

\begin{description}
\item[Planets in the globally mixed status]  
 If the envelope metallicity becomes comparable to the bulk one, 
 then mixing of heavy elements should take place across different layers (i.e., the gases accreted at different stages).
 Such mixing does not necessarily lead to the well-mixed status.
 Solar system giants are very likely classified into this type. 
\item[Planets in the locally (well-)mixed status]
 If the envelope metallicity is well reproduced by the onion-like model, 
 then mixing of heavy elements in the outermost layer should be effective, but mixing across different layers should be minimized.
 Explicit identification of exoplanets in this status serves as direct evidence of sequential accretion of gas and solids and supports the usefulness of the onion-like model.
\end{description}

The new classification of planets provides novel implications for planet formation.
For planets in the globally mixed status, mixing between different layers would occur during and/or after the epoch of planet formation and should be maintained in the present day.
On the other hand, for those in the locally (well-)mixed status, local mixing established during the epoch of planet formation should be maintained all the time.
If all planets would undergo the locally (well-)mixed status at the end of the planet formation epoch,
then a transition from the locally (well-)mixed status to the globally mixed one might be explained by dynamical processes such as giant impacts.
For solar system planets, giant impacts are viewed as one important process after the epoch of planet formation in protoplanetary disks;
the diluted core of Jupiter can be produced by a giant impact \citep{2019Natur.572..355L},
which would dredge up interior materials and could enhance mixing of heavy elements,
and the obliquity of Uranus and satellite formation around it can be explained by a giant impact as well \citep{2020NatAs...4..880I}.

The presence of two mixed statuses might be relevant to two different formation channels.
For the origin of hot Jupiters, both disk-induced and high-eccentricity migration are proposed \citep[][]{2018ARA&A..56..175D}.
The former hypothesis assumes that the formation process would be completed in gas disks,
while for the latter, hot-Jupiters would be exposed to residual planetesimals during a phase of eccentricity damping in the post gas disk era.
If planets accrete these planetesimals additionally during the phase, planets formed by the latter might end up in the globally mixed status.
Those formed by the former might remain in the locally (well-)mixed status.
Two formation channels are also proposed for sub-Neptune mass planets \citep[e.g.,][]{2019PNAS..116.9723Z,2023ApJ...947L..19R}:
the one leads to planets with tenuous H/He atmospheres, while the other results in water worlds.
The former planets might reside in the locally (well-)mixed status due to the presence of atmospheres,
and the latter might be classified into the globally mixed status due to the high abundance of volatile.
 
A potential complication that we identify as an important topic for further study is the possibility of rain-out (i.e., supersaturation) or settling of heavy elements \citep{2009A&A...507.1671M}. 
If the envelope metallicity is lower than the bulk one and/or deviates from the prediction of the onion-like model 
(see Figure \ref{fig4} in Appendix \ref{sec:app_int} for an example of a primordial interior profile of heavy elements), 
then heavy elements in the outermost layer rain-out. 
Such outliers exist in Figure \ref{fig3}, that is, planets are not covered by the shaped ellipses.
For these planets, rain-out would occur during and/or after the epoch of planet formation and could wash out the formation history of planets.

We admit that there are a number of caveats in this work.
First, this work is motivated by an expectation that the composition of the top thin layer of exoplanet atmospheres probed by transit observations should be relevant to that of deep interiors.
This is not verified in the literature yet.
Second, the origin of both the mixed statuses remains to be addressed. 
It might be related to the composition of accreted solids and their accretion efficiency, 
both of which would eventually regulate the envelope evolution (e.g., the excitation and sustainability of convention in planet interior and/or (non-)equilibrium chemical evolution).
However, the validity of the onion-like model could be challenged critically 
if detailed properties (e.g., radiative and convective zones) of atmospheres and relevant dynamical and chemical processes (e.g., condensation) are considered.
For instance, a recent detailed model suggests the importance of rain-out \citep[e.g.,][]{2022PSJ.....3...74S}, which is a warning against the onion-like model.
Since rain-out arises when the partial pressure of certain vapors exceeds their vapor pressure,
the composition of accreted solids and chemical reactions that possibly occur after thermal ablation and vaporization of certain materials would be a key to examine the validity of the onion-like model.
Third, our calculations are very simple, while the adopted models are based on detailed calculations and simulations (Appendix \ref{sec:app_model}).
In the literature, a similar onion-like model is proposed, focusing on the formation of Jupiter \citep{2017ApJ...840L...4H}.
In summary, it is important to determine under what conditions, our model and assumptions would be verified.

The advent of JWST rapidly improves the quality and volume of observational data and significantly accelerates exoplanet characterization \citep[e.g.,][]{2024RvMG...90..411K}.
While systematic analysis of JWST data remains to be conducted,
the exoplanet classification proposed in this work will be further tested by the analysis and used to advance our understanding of planet formation.
This work therefore provides one key reference of how both the bulk and atmospheric metallicities can be used to constrain how planets accrete gas and solids from their natal protoplanetary disks.

\begin{acknowledgments}

The authors thank an anonymous referee for useful comments on our manuscript.
This research was carried out at the Jet Propulsion Laboratory, California Institute of Technology, 
under a contract with the National Aeronautics and Space Administration (80NM0018D0004).
Y.H. is supported by JPL/Caltech and the NASA Exoplanets Research Program through grant 20-XRP20 2-0008.

\end{acknowledgments}

\appendix

\section{Formulation of planetary growth by core accretion} \label{sec:app_model}

We adopt the standard picture of core accretion for planet formation \citep{1996Icar..124...62P,2000Icar..143....2B,2012A&A...547A.111M},
wherein four stages are considered (Figure \ref{fig1}). 
As shown in Section \ref{sec:comp}, planetesimal accretion can better reproduce the profiles of the bulk and envelope metallicities of observed exoplanets.
An additional discussion is also provided below (Table \ref{table1}).
Most of physical quantities are defined in the main text.

\subsection{Gas accretion}

The main focus of this work is on Stages II to IV as these stages determine atmospheric metallicities of planets predominantly.

{\it Stage II:} Gas accretion onto (proto)planets becomes possible when gaseous envelopes around planetary cores cannot achieve a hydrostatic configuration
\citep{1980PThPh..64..544M,1982P&SS...30..755S,1986Icar...67..391B}.
The envelopes then contract gravitationally.
The corresponding Kelvin-Helmholtz timescale is given as \citep{2000ApJ...537.1013I}
\begin{equation}
\label{eq:tau_KH}
\tau_{\rm KH} = 10^{c} f_{\rm grain} \left( \frac{M_{\rm p}}{ 10 M_{\oplus} } \right)^{-d} \mbox{yr},
\end{equation}
where we set that $c=7$ and $d=4$, following \citet{1997Icar..126..282T}.
The resulting gas accretion rate ($\dot{M}_{XY, \rm KH}$) onto (proto)planets is written as
\begin{equation}
\label{eq:Mpdot_KH}
\dot{M}_{XY, \rm KH}  \simeq         \frac{M_{\rm p}}{ \tau_{\rm KH} } 
                                = 10^{-3}  \left( \frac{f_{\rm grain}}{ 10^{-3} } \right)^{-1}  \left( \frac{M_{\rm p}}{ 10 M_{\oplus} } \right)^{5} \frac{M_{\oplus}}{\mbox{yr}}.
\end{equation}
As described in Section \ref{sec:pred}, $f_{\rm grain}$ is treated as a parameter in this work 
since its dependence may not be crucial due to the power of 1/5 (see equation (\ref{eq:Zenv_StII})).
In reality, however, it is a function of grain properties (e.g., size distribution and composition),
and a detailed treatment is optimal, wherein the value of $f_{\rm grain}$ directly reflects the composition of accreted solids and their subsequent evolution.

{\it Stage III:} Spherically symmetric gas accretion (i.e., Stage II) ends when gas supply from the surrounding disk cannot catch up with the value of $\dot{M}_{XY, \rm KH}$.
The resulting gas accretion rate is regulated by disk evolution and written as \citep{2007ApJ...667..557T}
\begin{eqnarray}
\label{eq:Mpdot_hydro}
\dot{M}_{XY, \rm hydro} & =          &  0.29 \left( \frac{H_{\rm g}}{r_{\rm p}} \right)^{-2}  \left( \frac{M_{\rm p}}{M_*} \right)^{4/3} \Sigma_{\rm g} r_{\rm p}^2 \Omega  \\ \nonumber
                                   & \simeq &  8.6 \times 10^{-3}     \left( \frac{\Sigma_{\rm g}}{1.5 \times 10^2 \mbox{ g cm}^{-2}}  \right)    \\ \nonumber
                                   & \times  &  \left( \frac{H_{\rm g}/r_{\rm p}}{0.05} \right)^{-2} \left( \frac{r_{\rm p} }{5 \mbox{ au}} \right)^{1/2}    \left( \frac{M_{\rm p}}{10 M_{\oplus}} \right)^{4/3}  
                                                      \frac{M_{\oplus}}{\mbox{yr}},
\end{eqnarray}
where $M_*=M_{\odot}$ is the mass of the central star, and $\Omega(=\sqrt{GM_*/r^3_{\rm p}})$ is the angular frequency.

{\it Stage IV:} As the planet mass increases due to gas accretion, 
planet-disk interactions start changing the surrounding gas surface density and eventually open up a gap \citep{2012ARA&A..50..211K}.
Then the resulting $\dot{M}_{XY, \rm gap} $ is given as \citep{2016ApJ...823...48T}
\begin{eqnarray}
\label{eq:Mpdot_gap}
\dot{M}_{XY, \rm gap} & =          & 8.5 \left( \frac{H_{\rm g}}{r_{\rm p}} \right) \left( \frac{M_{\rm p}}{M_*} \right)^{-2/3}  \Sigma_{\rm g} \nu \\ \nonumber
                                 & \simeq & 1.9 \times 10^{-1}  \left( \frac{\alpha}{10^{-2}}  \right)\left( \frac{\Sigma_{\rm g}}{1.5 \times 10^2 \mbox{ g cm}^{-2}}  \right)     \\ \nonumber
                                 & \times  &   \left( \frac{H_{\rm g}/r_{\rm p}}{0.05} \right)^3  \left( \frac{r_{\rm p} }{5 \mbox{ au}} \right)^{1/2}  
                                                     \left( \frac{M_{\rm p}}{100M_{\oplus}} \right)^{-2/3}    \frac{M_{\oplus}}{\mbox{yr}},
\end{eqnarray}
where $\nu=\alpha H_{\rm g} / \Omega$ is the gas viscosity.

\subsection{Solid accretion} 

We now consider solid accretion.
We focus on three characteristic solids that are classified by size: dust, pebbles, and planetesimals.
Dust is typically $\lesssim 1$ mm in size and coupled well with gas.
Pebbles are from a few cm to even a few m in size, depending on the gas surface density.
They are large enough to de-couple from gas.
The resulting radial drift can expand an effective accretion zone of growing (proto)planets,
and hence pebbles are currently viewed as one important agent for core formation as well as possibly metal enrichment of giant planets.
Planetesimals are the largest objects with an order of a few km to a few hundred km in size.
They are massive enough that gravitational interactions with accreting objects regulate their dynamics predominantly.
Historically, planetesimals are regarded as the most critical body for metal enrichment.
We here explore solid accretion of these objects separately.

{\it Dust accretion:} As described, dust is well coupled with gas.
Hence, the resulting accretion rate ($\dot{M}_{Z, \rm dust}$) can be written as
\begin{equation}
\dot{M}_{Z, \rm dust} \simeq Z_{XY} \dot{M}_{XY},
\end{equation}
where $Z_{XY}$ is the metallicity of accreted gas, 
and for simplicity, it is here assumed to be the same as the host stellar metallicity ($Z_*$), that is $Z_{XY} \simeq Z_* \simeq 0.01$.

{\it Pebble accretion:} Pebble accretion becomes effective when the dynamics of solids is not approximated well by a Keplerian orbit.
This occurs for solids with the Stokes number of $\lesssim 2$ \citep{2012ApJ...747..115O}.
Previous studies formulate the resulting accretion rate \citep{2012ApJ...747..115O,2022ApJ...935..101H}, 
and we find that the rate takes a peak value around the Stokes number of $\sim 0.7$.
Then the simplified expression can be written as
\begin{eqnarray}
\dot{M}_{Z, \rm peb}  & \simeq & 3.6 \Sigma_{\rm s} R_{\rm H}^2 \Omega \\ \nonumber
                                  & \simeq & 2.4 \times 10^{-3}  \left( \frac{\Sigma_{\rm s}}{2.7 \mbox{ g cm}^{-2}}  \right)    \\ \nonumber
                                 & \times  &  \left( \frac{r_{\rm p} }{5 \mbox{ au}} \right)^{1/2}   \left( \frac{M_{\rm p}}{10M_{\oplus}} \right)^{2/3}      \frac{M_{\oplus}}{\mbox{yr}},
\end{eqnarray}
where the numerical factor is estimated at the Stokes number of 0.7, 
and $R_{\rm H} (=r_{\rm p} [M_{\rm p}/(3M_*)]^{1/3})$ is the Hill radius of accreting (proto)planets.
The above expression is applicable for the so-called Hill regime \citep{2012A&A...544A..32L}, 
and we confirm that the condition holds for pebbles with the Stokes number of $ \sim 0.7$.

{\it Planetesimal accretion:} The accretion rate of planetesimals onto (proto)planets depends on the presence or absence of gaps in planetesimal disks \citep{2008ApJ...684.1416S}.
The corresponding accretion rates are written as, for the case of no gap formation 
\begin{eqnarray}
\dot{M}_{Z, \rm nogap} & \simeq &2.2 \times 10^{-6}  \left( \frac{\dot{M}_{XY}}{10^{-4} M_{\oplus}\mbox{ yr}^{-1}} \right)^{4/5}     \\ \nonumber
                                       & \times &\left( \frac{\Sigma_{\rm s}}{2.7 \mbox{ g cm}^{-2}}  \right)  \left( \frac{r_{\rm p}}{5 \mbox{ au}} \right)^{6/5}   \\ \nonumber
                                       & \times & \left( \frac{\rho}{1 \mbox{ g cm}^{-3}} \right)^{1/2}  \left( \frac{R_{\rm p}}{R_{\oplus}} \right)^2  \left( \frac{M_{\rm p}}{M_{\oplus}} \right)^{-16/15}   
                                                          \frac{M_{\oplus}}{\mbox{yr}},
\end{eqnarray}
and, for the case of gap formation 
\begin{eqnarray}
\dot{M}_{Z, \rm gap} & \simeq &5.2 \times 10^{-7}  \left( \frac{\tau_{\rm damp}}{10^4 \mbox{ yr}} \right)^{7/10}  \left( \frac{\dot{M}_{XY}}{10^{-4} M_{\oplus}\mbox{ yr}^{-1}} \right)^{7/5}    \\ \nonumber
                                       & \times & \left( \frac{\Sigma_{\rm s}}{2.7 \mbox{ g cm}^{-2}}  \right)  \left( \frac{r_{\rm p}}{5 \mbox{ au}} \right)^{21/20}   \\ \nonumber
                                       & \times & \left( \frac{\rho}{1 \mbox{ g cm}^{-3}} \right)^{1/2}  \left( \frac{R_{\rm p}}{R_{\oplus}} \right)^2  \left( \frac{M_{\rm p}}{M_{\oplus}} \right)^{-49/30}  
                                                         \frac{M_{\oplus}}{\mbox{yr}}.
\end{eqnarray}

The above equations exhibit explicit dependence of $\dot{M}_{Z}$ on $\dot{M}_{XY}$ and $R_{\rm p}$, 
suggesting that the resulting accretion rate varies when accreting planets undergo different gas accretion modes and/or different mass-radius relations.
Such dependence becomes important when exploring how bulk and atmospheric metallicities can serve as direct probes of sequentially changing accretion modes.

\subsection{Mass-radius relation}

We formulate  mass-radius relations used in this work.
Since we consider the formation stages, an {\it effective} mass-radius relation is adopted.

For Stage II,
planets undergo spherically symmetric accretion.
In such a case, the envelope of planets extends to their Hill radius and they are fully attached to their natal protoplanetary disks.
The effective radius of solid accretion depends sensitively on the size of accreted solids.
Previous studies show that $\gtrsim$ 100 km-sized bodies are accreted almost at the surface of planetary cores \citep{2003A&A...410..711I}.
Hence, we adopt the following:
\begin{equation}
R_{\rm p} \simeq 1R_{\oplus} ( M_{\rm p} / M_{\oplus} )^{1/3}.
\end{equation}

For Stages III and IV, spherically symmetric accretion ends, and gas accretion onto planets is regulated by disk evolution.
In such a case, planetary envelopes should shrink from the Hill radius \citep{2000Icar..143....2B,2012A&A...547A.111M}, 
and their size would become broadly comparable to the size of mature planets.
We therefore adopt the mass-radius relation derived from observed exoplanets \citep{2017ApJ...834...17C}.

In summary, we use the following relationship:
\begin{equation}
R_{\rm p} \simeq  \left\{ 
                           \begin{array} {l}
                           		       {1R_{\oplus} ( M_{\rm p} / M_{\oplus} )^{1/3}   }              \ \ \     (4 \lesssim M_{\rm p}/M_{\oplus} \lesssim 18), \\
                                                 {1.2 R_{\oplus} ( M_{\rm p} / 2M_{\oplus} )^{3/5}   }          \ \ \    (18 \lesssim M_{\rm p}/M_{\oplus} \lesssim 10^2), \\
                                                 {12 R_{\oplus} \simeq1.1 R_{\rm Jup}  }                    \ \ \  (10^2 \lesssim M_{\rm p}/M_{\oplus} \lesssim 10^4). 
                           \end{array} 
                     \right.
\end{equation}
Equivalently, 
\begin{equation}
\rho \simeq  \left\{ 
                           \begin{array} {l}
                                                 {\rho_{\oplus}   }                                                             \ \ \           (4 \lesssim M_{\rm p}/M_{\oplus} \lesssim 18), \\
                                                 {\rho_{\oplus}(2^{3/5}/1.2)^3 ( M_{\rm p} / M_{\oplus} )^{-4/5}   }       \ \ \             (18 \lesssim M_{\rm p}/M_{\oplus} \lesssim 10^2), \\
                                                 { \rho_{\oplus}/12^3 ( M_{\rm p} / M_{\oplus} ) }                      \ \ \    (10^2 \lesssim M_{\rm p}/M_{\oplus} \lesssim 10^4), 
                           \end{array} 
                     \right.
\end{equation}
where $\rho_{\oplus}=5$ g cm$^{-3}$ is the bulk density of the Earth.

\subsection{Power-law indices of the bulk and envelope metallicities}

We now combine all the components and compute the profiles of bulk and envelope metallicities.

As described in Section \ref{sec:comp}, we focus on the profile and confirm the finding of previous studies \citep{2018ApJ...865...32H};
planetesimal accretion provides a most consistent trend for the total heavy-element mass of observed exoplanets \citep[i.e., $Z_{\rm p, bulk} \simeq -0.45 \pm0.09$ from ][]{2016ApJ...831...64T};
the slopes predicted from dust accretion are zero in Stages II to IV, and those from pebble accretion are zero in Stage II and too steep in Stages III and IV (Table \ref{table1}).

For envelope metallicities, a dedicated fit is not conducted, as two populations (high and low envelope metallicities) may exist (Figure \ref{fig2}).
However, planetesimal accretion clearly gives a most reasonable explanation, as shown in Figure \ref{fig2} (also see Table \ref{table1}).

Thus, it can be concluded that planetesimal accretion is a most promising process to reproduce the bulk and atmospheric metallicity profiles of currently observed exoplanets.

\begin{table*}
\begin{minipage}{17cm}
\centering
\caption{The computed power-law indices of the bulk and envelope metallicities}
\label{table1}
{
\begin{tabular}{c|c|c|c}
\hline \hline
                     &  Stage II                                                            &  Stage III                                                                &  Stage IV          \\ \hline 
Planet Mass & $ 4 \lesssim M_{\rm p}/M_{\oplus} \lesssim 18 $      &  $ 18 \lesssim M_{\rm p}/M_{\oplus} \lesssim 10^2 $       &  $ 10^2 \lesssim M_{\rm p}/M_{\oplus} \lesssim 10^4 $  \\ \hline \hline
Expression  & $Z_{\rm p, bulk} \simeq 1$                               & $Z_{\rm p, bulk} \simeq \dot{M}_{Z, \rm env}  / \dot{M}_{XY, \rm hyrdo} $   & $Z_{\rm p, bulk} \simeq \dot{M}_{Z, \rm env}  / \dot{M}_{XY, \rm hydro}$  \\ 
                    & $Z_{\rm p, env} \simeq  \dot{M}_{Z, \rm env}  / \dot{M}_{XY, \rm KH} $ &  $Z_{\rm p, env} \simeq \dot{M}_{Z, \rm env}  / \dot{M}_{XY, \rm hydro}$   &  $Z_{\rm p, env} \simeq \dot{M}_{Z, \rm env}  / \dot{M}_{XY, \rm gap}$ \\ \hline \hline
Planetesimal & $Z_{\rm p, bulk} \simeq 1$                               & ${\bf Z_{\rm p, bulk} \propto M_{\rm p}^{-8/15}}$                      & ${\bf Z_{\rm p, bulk} \propto M_{\rm p}^{-8/15}}$           \\
w/o gaps        &  ${\bf Z_{\rm p, env} \propto M_{\rm p}^{-7/5}}$       &  ${\bf Z_{\rm p, env} \propto M_{\rm p}^{-8/15}}$               & $ Z_{\rm p, env} \propto M_{\rm p}^{-13/30}$ \\ \hline  
Planetesimal & $Z_{\rm p, bulk} \simeq 1$                               & $ {\bf Z_{\rm p, bulk} \propto M_{\rm p}^{-3/10}}$                      & $ {\bf Z_{\rm p, bulk} \propto M_{\rm p}^{-3/10}}$           \\
w/ gaps         &  $Z_{\rm p, env} \propto M_{\rm p}^{31/30}$            &  ${\bf Z_{\rm p, env} \propto M_{\rm p}^{-3/10}}$                & ${\bf Z_{\rm p, env} \propto M_{\rm p}^{-7/5}}$ \\ \hline  \hline
Pebble          & $Z_{\rm p, bulk} \simeq 1$                               & $Z_{\rm p, bulk} \propto M_{\rm p}^{-2/3}$                          & $Z_{\rm p, bulk} \propto M_{\rm p}^{-2/3}$           \\
                    &  $Z_{\rm p, env} \propto M_{\rm p}^{-13/3}$             &  $Z_{\rm p, env} \propto M_{\rm p}^{-2/3}$                          & $Z_{\rm p, env} \propto M_{\rm p}^{4/3}$ \\ \hline \hline
Dust             & $Z_{\rm p, bulk} \simeq 1$                               & $Z_{\rm p, bulk} \simeq 10^{-2}$                                & $Z_{\rm p, bulk} \simeq 10^{-2}$           \\
                    &  $Z_{\rm p, env} \simeq 10^{-2}$                      &  $Z_{\rm p, env} \simeq 10^{-2}$                                & $Z_{\rm p, env} \simeq 10^{-2}$ \\ \hline 
\hline              
\end{tabular}

}
\end{minipage}
\end{table*}

\section{Exoplanet data} \label{sec:app_data}

We make use of two kinds of exoplanet data available in the literature.
The first kind are bulk metallicities of exoplanets.
These data are computed, by combining precise measurements of planet mass and radius with thermal evolution of planets \citep{2016ApJ...831...64T}.
Since the radius change for a given mass of planets is controlled predominantly by the amount of heavy elements contained in the planets,
precise measurements of the current planet size can be used to infer the total heavy elements mass of planets.
Specifically, 47 exoplanets are selected from larger samples and their heavy element masses are computed,
which are summarized in table 1 of \citet{2016ApJ...831...64T}.
We adopt these values from the table.

The second kind of the data are envelope metallicities of exoplanets.
Currently, transit observations are leveraged to compute the enhancement/reduction factor of elements residing in exoplanet atmospheres.
Conducting retrieval analysis, the abundance-based, atmospheric metallicities are obtained.
We adopt such metallicities  from \citet{2023ApJS..269...31E}.

In order to directly compare the atmospheric metallicity with the bulk metallicity,
we convert the abundance-based, atmospheric metallicities to the mass-based one as follows \citep[also see][]{2024SSRv..220...61S}.

Suppose that retrieval analysis infers the abundance enhancement/reduction (X/H) of an element (X) relative to solar abundance in an exoplanet atmosphere.
Then, the resulting mass fraction ($m_{\rm X}$) of the element is written as
\begin{equation}
    m_{\rm X} = A_{\rm X} N_{\rm X} \frac{\rm X}{\rm H},
\end{equation}
where $A_{\rm X}$ and $N_{\rm X}$ are the nucleons and the number density of the element at solar metallicity, respectively, 
and $m_{\rm X}$ is normalized by the atomic mass unit.
The values of $N_{\rm X}$ are adopted from \citet{2009ARA&A..47..481A}.

If the C/O ratio is also determined, which tends to be common for recent retrieval analyses, 
then the total number ($N_{\rm tot}$) of elements in the atmosphere is given as
\begin{equation}
    N_{\rm tot} = \sum_{\rm X}  f_{\rm C/O}  N_{\rm X} \frac{\rm X}{\rm H},
\end{equation}
where it is assumed that X/H = 1 for hydrogen and helium, and the C/O ratio is defined as:
\begin{equation}
    f_{\rm C/O} = \left\{ 
                           \begin{array} {l}
                                                 {1} \ ({\rm X} \neq {\rm C}), \\
                                                  {\rm C/O} \ ({\rm X} = {\rm C}). 
                           \end{array} 
                     \right.
\end{equation}
In the above calculation, the solar value is adopted for $N_{\rm H_e}$, and $N_{\rm H}$ is constrained such that $N_{\rm tot}$ becomes unity.

Consequently, the mass-based, atmospheric metallicity is computed as
\begin{equation}
    \label{eq:Zp_env_ob}
    Z_{\rm p, env}^{\rm ob} \equiv \sum_{{\rm X} \neq {\rm H, He}} f_{\rm C/O} A_{\rm X} N_{\rm X} \frac{\rm X}{\rm H} \ /  \ \sum_{{\rm X} } f_{\rm C/O} A_{\rm X} N_{\rm X} \frac{\rm X}{\rm H}.
\end{equation}

In equation (\ref{eq:Zp_env_ob}), one value of X/H is used for all the elements to compute $Z_{\rm p, env}^{\rm ob}$ unless the C/O ratio is specified.
This is motivated by the fact that except for hydrogen and helium, 
the composition of meteorites in the solar system is broadly comparable to that of the sun \citep[e.g.,][]{1989GeCoA..53..197A}.
In other words, this approach implicitly assumes that the composition of accreted solids be comparable to that of the sun (except for hydrogen and helium)
and the enhancement/reduction of X/H directly reflects the number of these solids accreted by planets,
which may also be relevant to the metallicity difference between the host star and the sun.
The mass-based metallicity is thus preferred when comparing with the envelope metallicity predicted from the onion-like model;
as described in Section \ref{sec:comp},
the computed mass-based metallicities would represent the {\it primordial} ones under the well-mixed assumption.

In reality, different solids may have different compositions, and material mixing occurs in planet interiors.
Hence, different elements would have different values of X/H.
However, current observations still do not enable reliable derivation of such values for all the elements for most exoplanets;
given that refractory materials turn into grains and sink towards planetary cores in their envelopes,
hot Jupiters would be the only targets that allow access to such detailed information \citep[e.g.,][]{2023Natur.620..292C}.
It should be noted that existing observations probe the water abundance predominantly \citep{2023ApJS..269...31E}. 
Since water is volatile, it would have higher chance to remain well-mixed ever since the formation stage.

Armed with the above formulation, we compute $Z_{\rm p, env}^{\rm ob}$ for 70 exoplanets.
In \citet{2023ApJS..269...31E}, a different expression (i.e., $Z_{\rm p}$) for X/H is used,
and we assume that $Z_{\rm p} \equiv$ X/H, as it is the standard outcome of retrieved analysis. 

\section{A primordial Interior profile of heavy elements} \label{sec:app_int}

\begin{figure}
\centering
\includegraphics[width=0.9\linewidth]{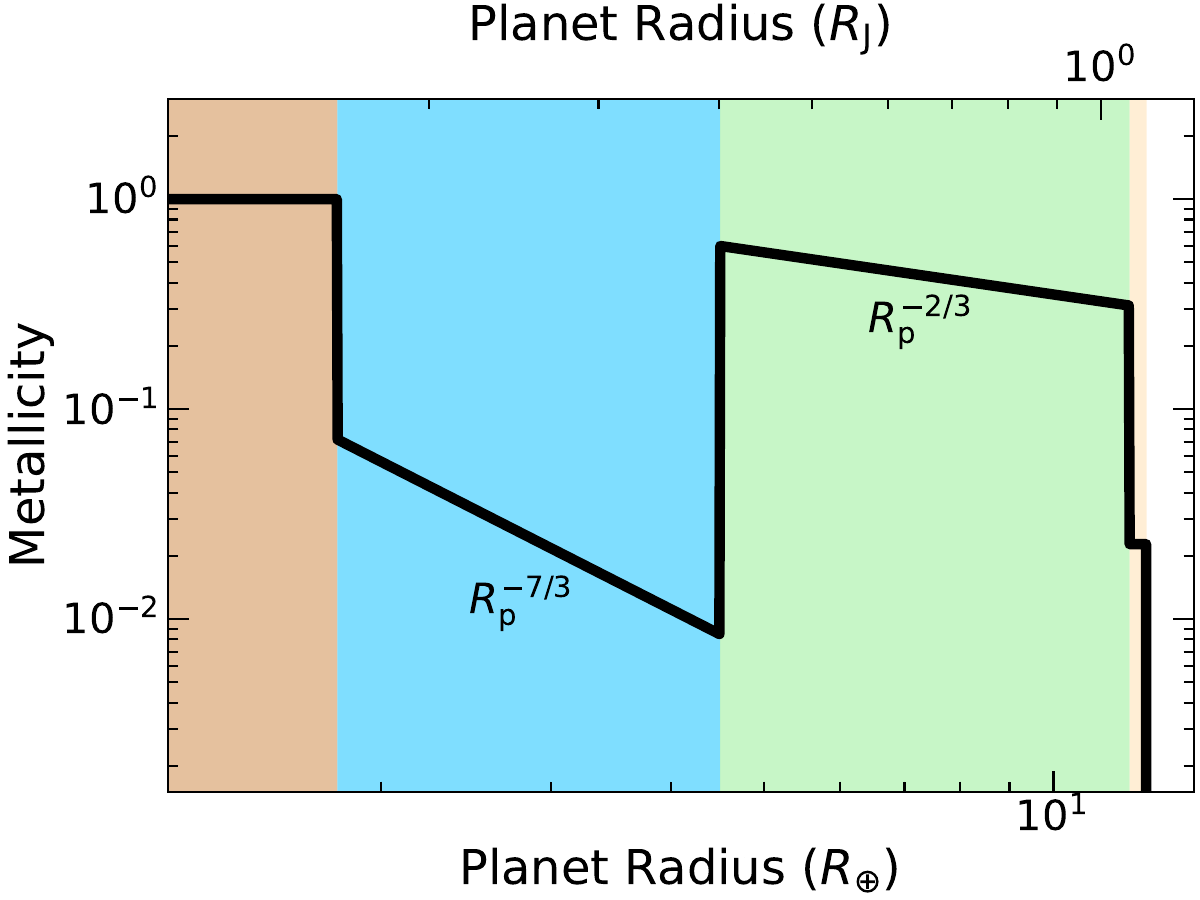}
\caption{An example of primordial interior profiles of heavy elements for the locally well-mixed case.
The profile changes from $Z_{\rm p} =1$ to $Z_{\rm p} \propto R_{\rm p}^{-7/3}$ and up to $Z_{\rm p} \propto R_{\rm p}^{-2/3}$.
The metallicity profile at planetary surfaces cannot be constrained by the current model due to the constant mass-radius relation there.}
\label{fig4}
\end{figure}

The onion-like model enables determination of primordial interior profiles of heavy elements.
As an example, a certain profile is plotted in Figure \ref{fig4}.
The normalization constant at each layer is derived from the envelope metallicities of observed exoplanets (see Figure \ref{fig2}),
and hence it should vary for different planets.
Also, the following mass-radius relation is used to convert the planet mass to radius:
\begin{equation}
M_{\rm p} \simeq  \left\{ 
                           \begin{array} {l}
                                                 {2 M_{\oplus} ( R_{\rm p} / 1.2R_{\oplus} )^{5/3}   }          \ \ \    (1.2 \lesssim R_{\rm p}/R_{\oplus} \lesssim 12), \\
                                                 {M_{\rm p}  }                    \ \ \  (12 \lesssim R_{\rm p}/R_{\oplus} ). 
                           \end{array} 
                     \right.
\end{equation}
Note that the planet radius becomes independent of the planet mass at Stage IV, leading to the constant profile at the stage.

The profile gives one reference and can be compared with more detailed models in which rain-out of heavy elements is taken into account \citep{2022PSJ.....3...74S}.
It can also be coupled with retrieval analysis of transit observations to constrain the metal distribution within observed exoplanets.

\bibliographystyle{aasjournal}
\bibliography{adsbibliography}    



\end{document}